\documentclass[prd,superscriptaddress,amsfonts,amssymb,amsmath,showpacs,twocolumn]{revtex4}
\usepackage{epsfig}
\usepackage[applemac]{inputenc}
\usepackage{graphicx}
\usepackage{hyperref}
\date{\today}

\begin{document}

\title{Particle creation due to tachyonic instability in relativistic stars}
\author{A.\ G.\ S.\ Landulfo}\email{andre.landulfo@ufabc.edu.br}
\affiliation{Instituto de F\'\i sica de S\~ao Carlos,
Universidade de S\~ao Paulo, Caixa Postal 369, CEP 13560-970, 
S\~ao Carlos, SP, Brazil}
\author{W.\ C.\ C.\ Lima}\email{wccl@ift.unesp.br}
\affiliation{Instituto de F\'\i sica de S\~ao Carlos,
Universidade de S\~ao Paulo, Caixa Postal 369, CEP 13560-970, 
S\~ao Carlos, SP, Brazil}
\author{G.\ E.\ A.\ Matsas}\email{matsas@ift.unesp.br}
\affiliation{Instituto de F\'\i sica Te\'orica, Universidade 
Estadual Paulista,
Rua Dr.\ Bento Teobaldo Ferraz, 271 - Bl.\ II, CEP 01140-070, 
S\~ao Paulo, SP, Brazil}
\author{D.\ A.\ T.\ Vanzella}\email{vanzella@ifsc.usp.br}
\affiliation{Instituto de F\'\i sica de S\~ao Carlos,
Universidade de S\~ao Paulo, Caixa Postal 369, CEP 13560-970, 
S\~ao Carlos, SP, Brazil}


\begin{abstract}

Dense enough compact objects were recently shown to lead to an exponentially 
fast increase of the vacuum energy density for some free scalar fields 
properly coupled to the spacetime curvature as a consequence of a tachyonic-like 
instability. Once the effect is triggered, the star energy density would be 
overwhelmed by the vacuum energy density in a few milliseconds. This demands that  
eventually geometry and field evolve to a new configuration to bring the vacuum 
back to a stationary regime. Here, we show that the vacuum fluctuations built 
up during the unstable epoch lead to particle creation in the final stationary 
state when the tachyonic instability ceases. The amount of created particles 
depends mostly on the duration of the unstable epoch and final stationary 
configuration, which  are open issues at this point.  We emphasize that the 
particle creation coming from the tachyonic instability will occur even in 
the adiabatic limit, where the spacetime geometry changes arbitrarily slowly, 
and therefore is quite distinct from the usual particle creation due to the 
change in the background geometry.
 
\end{abstract}

\maketitle

\section{Introduction}

It was recently shown that  relativistic stars may become unstable due 
to quantum-field effects~\cite{LV10,LMV10}. The so-called {\em vacuum awakening 
effect} occurs for a free scalar field $\Phi$ properly coupled to the 
spacetime curvature~\cite{LV10}. This effect is characterized by an 
exponential point-dependent increase and decrease of the vacuum 
expectation value of the stress-energy-momentum tensor $\langle \hat 
T_{a b} \rangle$. This is caused by a tachyonic-like instability,
which induces an exponential growth of $\langle {\hat \Phi}^2 \rangle$. Once the 
effect is triggered and the scalar field exits its initial quiescent regime, 
few milliseconds would be enough for the vacuum to overwhelm the 
energy density of the compact object. The star destiny is presently 
uncertain because it depends on  how scalar field 
and spacetime geometry evolve in the unstable phase to reach a 
final stable configuration. As recently argued in Ref.~\cite{PCBRS11}, in some cases
the appearance of a proper scalar field could restabilize the star, a phenomenon 
usually called {\em spontaneous scalarization}~\cite{DE93}. This would typically change 
the star gravitational mass by a few percent. However, depending on how the 
star enters the unstable phase, it seems possible that the scalar field does 
not react fast enough, leading to some dramatic implosion/explosion event.
Whatever turns out to be the final configuration, being the star somehow 
rebalanced or destroyed, the unstable phase must be detained and the vacuum 
must evolve to some new stationary regime. This observation alone 
allows us to extract quite important information about the final state of the 
scalar field as the vacuum  ``falls asleep" again. In particular,  we show that 
the vacuum fluctuation built up during the unstable epoch leads to particle 
creation in the final stationary state. The amount of created particles 
will depend mostly on the duration of the unstable epoch and final stationary
configuration.

The paper is organized as follows. In Sec.~\ref{field_quantization}, we discuss 
the quantization procedure for a free scalar field in a curved spacetime 
in the presence of tachyonic-like modes. In Sec.~\ref{awaking_the_vacuum_in_stars}, 
we apply the previous-section results to review the vacuum awakening effect 
in relativistic stars. In Sec.~\ref{detectors}, we probe the unstable 
phase using Unruh-DeWitt detectors. In this period, no natural particle content 
can be ascribed to the scalar field and the use of detectors is particularly useful 
to investigate the behavior of the vacuum fluctuation. We show in this section that even 
assuming a static spacetime in the unstable phase when the vacuum 
is ``awake", particle detectors following orbits of the timelike isometry will 
copiously excite. This is possible according to co-static observers because 
each detector excitation is accompanied by a corresponding decrease in the energy 
stored in the field due to the excitation of a nonstationary (tachyonic) mode which 
contributes with negative energy. Then, in Sec.~\ref{particle creation}, we show 
that  at least part of the quantum fluctuations built up in the awaken phase 
eventually draws to particle creation as the unstable period ends and the vacuum falls 
dormant again. We emphasize that the particle creation occurs even 
assuming that the spacetime change is arbitrarily 
slow and, thus, differs, e.g., from the well-known phenomenon of particle creation 
in evolving universes induced by the change of the background geometry~\cite{Parker68, 
Parker69, BD82}. A toy model is also offered to illustrate 
in a concrete scenario the conclusions above. We close 
the paper with our final remarks in Sec.~\ref{final remarks}. 

\section{Preliminaries: free scalar field quantization with tachyonic-like 
modes}
\label{field_quantization}

\subsection{Standard field quantization in globally hyperbolic spacetimes}
\label{standard_quantization_general_spacetimes}

In this section, we review the standard quantization of free scalar fields 
in curved spaces~\cite{Wald94, Parker09} giving particular attention 
to the case of static spacetimes with tachyonic-like modes. Let  us begin 
by considering a globally hyperbolic spacetime $(\mathcal{M},g_{a b})$ 
foliated with Cauchy surfaces $\Sigma_t$ labeled with a parameter $t$. 
Now, let us cover the $\Sigma_t$ surfaces with ${x^i}$ ($i=1,2,3$) 
coordinates satisfying $n^a \nabla_a x^i =0$, where $n^a$ is the future-directed 
unit vector field normal to $\Sigma_t$. In terms of coordinates 
$x=(t, x^i)$, we can write the spacetime line element as
\begin{equation}
ds^2 = N^2 (-dt^2 + h_{i j} (x)\, dx^i dx^j),
\label{general line element}
\end{equation}
where $N=N(x)>0$ is the lapse function and $^{(3)}g_{ i j} = N^2 h_{i j}$ 
is the three-dimensional spatial metric induced on each Cauchy surface 
$\Sigma_t$.

We define the dynamics of a real scalar field $\Phi$ with mass $m$ in $(\mathcal{M},g_{a b})$ 
through the action
\begin{equation}\label{kg_action}
S \equiv 
-\frac{1}{2}
\int_{\mathcal{M}} d^4x \sqrt{-g} \,
( \nabla_a\Phi \nabla^a\Phi + m^2 \Phi^2 + \xi R \Phi^2 ),
\end{equation}
where $g \equiv {\rm det} (g_{ab})$ and $\xi \in \mathbb{R}$ determines the
non-minimal coupling between the scalar field and the scalar curvature
$R$. This leads to the following Klein-Gordon equation: 
\begin{equation}\label{kg_eq}
- \nabla_a\nabla^a\Phi + m^2\Phi + \xi R \Phi=0.
\end{equation}
Next, we define the Klein-Gordon inner product between any two solutions 
$u$ and $v$ of Eq.~(\ref{kg_eq}) as
\begin{equation}
(u,v)_{\rm KG} 
\equiv 
i \int_{\Sigma_t} d\Sigma \, n^a [ u^* \nabla_a v - v \nabla_a u^* ],
\label{inner_KG}
\end{equation}
where 
$d \Sigma$ is the proper-volume element on $\Sigma_t$,
and we recall that Eq.~(\ref{inner_KG}) does not depend on the 
choice of $\Sigma_t$. 

 The conjugate-momentum density $\Pi (x)$ is defined as
\begin{equation}\label{conjugate_canonical_momentum}
\Pi\equiv {\delta S}/{\delta \dot{\Phi}} = \sqrt{^{(3)} g\, }\, n^a\nabla_a\Phi,
\end{equation}
where $``\; \dot{}\;" \equiv \partial_t $ and 
$^{(3)} g \equiv {\rm det} (^{(3)}g_{i j})$.
The canonical quantization procedure consists of promoting field 
and momentum density to operators $\hat{\Phi}$ and $\hat{\Pi}$, 
respectively, satisfying canonical commutation relations:
\begin{eqnarray}
\, [\hat{\Phi} (t, {\bf x}), \hat{\Phi} (t, {\bf x}') ]_{\Sigma_t} 
& = & 
[\hat{\Pi}  (t, {\bf x}), \hat{\Pi}  (t, {\bf x}') ]_{\Sigma_t} = 0,
\label{canonical_quantization1} 
\\
\, [\hat \Phi (t, {\bf x}), \hat{\Pi}  (t, {\bf x}')]_{\Sigma_t}
& = &  
i \delta^3 ({\bf x}, {\bf x}' ),
\label{canonical_quantization2}
\end{eqnarray}
where ${\bf x} \equiv (x^1,x^2,x^3)$.

In order to realize a representation of these commutation relations, 
consider positive- and negative-norm solutions of Eq.~(\ref{kg_eq}),
$u_\alpha^{(+)}$ and $u_\alpha^{(-)} \equiv  (u_\alpha^{(+)})^*$, 
respectively, which together form a complete set of normal modes satisfying
\begin{eqnarray}
(u_\alpha^{(+)}, u_\beta^{(+)})_{\rm KG}
& = &
-(u_\alpha^{(-)}, u_\beta^{(-)})_{\rm KG}= \delta(\alpha, \beta),
\label{KGu1} 
\\
(u_\alpha^{(+)}, u_\beta^{(-)})_{\rm KG}
& = &
0.
\label{KGu2} 
\end{eqnarray}
Here, $\delta(\alpha, \beta)$ is the delta function
associated with the quantum numbers formally described by
$\alpha$, $\beta$. Then, we construct the field operator 
using $\{u_\sigma^{(+)}, u_\sigma^{(-)} \}$
as
\begin{equation}
\hat \Phi 
= 
\int d\mu(\sigma) 
[\hat a_\sigma u_\sigma^{(+)}+\hat a_\sigma^\dagger u_\sigma^{(-)}],
\label{Phi_expandido}
\end{equation} 
where $\mu$ is a measure defined on the set of quantum numbers
and in order to satisfy 
Eqs.~(\ref{canonical_quantization1}) and~(\ref{canonical_quantization2}),
the annihilation $\hat a_\sigma$ and creation $\hat a_\sigma^\dagger$
operators must satisfy the usual commutation relations
$[\hat a_\alpha, \hat a_\beta^\dagger] = \delta (\alpha, \beta)$,
$[\hat a_\alpha, \hat a_\beta] = 0$. The vacuum state $| 0 \rangle$
of this representation is defined by requiring $\hat a_\sigma |0\rangle = 0$ 
for all $\sigma$.    

\subsection{Quantum fields in static spacetimes}
\label{standard_quantization_static_spacetimes}

Now, we restrict our analysis to static spacetimes
in which case the line element~(\ref{general line element}) is cast as
\begin{equation}
ds^2 = N^2  (-dt^2 + h_{i j} ({\bf x})\, dx^i dx^j),
\label{static line element}
\end{equation}
where $N= N({\bf x}) >0$.
Under this condition, we write the field equation~(\ref{kg_eq}) in the form
\begin{equation}\label{kg_eq_static}
-\frac{\partial^2 \tilde \Phi}{\partial t^2 }= [- \Delta + V_{\rm eff}(x)] \tilde \Phi,
\end{equation}
where 
$\tilde\Phi \equiv N \Phi$,
$\Delta $ is the 
Laplace operator associated with $h_{ij}$, and
\begin{eqnarray}
V_{\rm eff} ({\bf x})
&=& N^{-1} \Delta N + N^2 (m^2 + \xi R)
\nonumber \\
&=& (1-6\xi ) N^{-1}  \Delta N + N^2 m^2 + \xi K
\end{eqnarray} 
is the effective potential with $K= K({\bf x})$ being the scalar curvature 
associated with $h_{ij}$.

The existence of a timelike Killing field   $\kappa^b = (\partial_t)^b$
associated with spacetime~(\ref{static line element}) suggests that 
we look for solutions of Eq.~(\ref{kg_eq_static}) in the form 
$\tilde u_\sigma^{(+)} \propto F_\sigma ({\bf x}) \exp (- i \omega_\sigma t)$
corresponding to solutions 
$
u_\sigma^{(+)} \propto N^{-1} F_\sigma ({\bf x}) 
\exp (- i \omega_\sigma t)
$
for Eq.~(\ref{kg_eq}). In this case, 
$F_\sigma ({\bf x})$ will satisfy
\begin{equation}
\left[-\Delta + V_{\rm eff}({\bf x})\right] 
F_{\sigma}({\bf x})
=\lambda_{\sigma}F_{\sigma}({\bf x}), 
\;\;\;  \lambda_\sigma = \omega_\sigma^2,
\label{av0}
\end{equation}
with proper boundary conditions. At this point, the only restriction
on $\lambda_\sigma$ is the one imposed by Hermiticity of the operator 
$-\Delta + V_{\rm eff}({\bf x})$, which demands 
$\lambda_\sigma \in \mathbb{Re}$.

Let us consider first solutions of 
Eq.~(\ref{av0}) with positive eigenvalues: $\lambda_\sigma 
\equiv \varpi_\sigma^2>0$. Then, the corresponding positive-norm 
solutions satisfying Eq.~(\ref{kg_eq}) will be the usual oscillatory 
modes: 
\begin{equation}
v^{(+)}_{\sigma}
= \frac{ e^{-i \varpi_{\sigma}t}}{\sqrt{2\varpi_{\sigma}\,} N({\bf x})}
F_{\sigma}({\bf x}), 
\;\;\; \varpi_{\sigma} > 0,
\label{intmodes0}
\end{equation}
where we demand
\begin{equation}
\int_{\Sigma_t} d^3 x \, \sqrt{h} \, F_\alpha({\bf x})^* F_\beta({\bf x}) 
= \delta (\alpha, \beta)
\label{demand}
\end{equation}
in order to guarantee that modes~(\ref{intmodes0}) are properly 
normalized according to Eqs.~(\ref{KGu1}) and~(\ref{KGu2}).

Now, we note that in some cases Eq.~(\ref{av0}) also allows for solutions with 
negative eigenvalues: $\lambda_\sigma \equiv -\Omega_\sigma^2<0$. These
solutions are associated with solutions of Eq.~(\ref{kg_eq_static}) with
exponentially increasing and decreasing $ \exp (\pm \Omega_\sigma t)$ time dependence. 
Under such circumstances, $\{v_\sigma^{(+)}, v_\sigma^{(-)} \}$ must be supplemented by 
an extra set of modes $\{w_\sigma^{(+)}, w_\sigma^{(-)} \}$ 
in order to generate a basis for the solution space of Eq.~(\ref{kg_eq}).  
Normalized positive-norm  modes $w_\sigma^{(+)}$ can be found and read 
\begin{equation}
w^{(+)}_{\sigma}
=
e^{i \beta_\sigma}
\frac{ 
\left(e^{ \Omega_{\sigma}t -i\alpha_\sigma}+ 
      e^{-\Omega_{\sigma}t +i\alpha_\sigma} 
\right)}
{\sqrt{4\Omega_{\sigma} \sin (2 \alpha_\sigma)} \,N({\bf x})
}F_{\sigma}({\bf x}),
\;\;\; \Omega_{\sigma} > 0,
\label{intmodes2a}
\end{equation}
where 
$\alpha_\sigma \in\ ]0,\frac{\pi}{4}]$ and, for the sake of 
convenience, we did not vanish the arbitrary global phase 
$\beta_\sigma$ yet. By choosing, e.g., 
$\alpha_\sigma = \beta_\sigma = \pi/4$, 
Eq.~(\ref{intmodes2a}) would assume the simple form
\begin{equation}
w^{(+)}_{\sigma} 
=
\frac{ 
\left(e^{ \Omega_{\sigma}t}+ 
      i e^{-\Omega_{\sigma}t } 
\right)}
{2 \sqrt{\Omega_{\sigma}} N({\bf x})  
}F_{\sigma}({\bf x}),
\;\;\; \Omega_{\sigma} > 0
\end{equation}
but we shall adopt here the same choice as in 
Ref.~\cite{LV10}, where $\beta_\sigma=0$ and $\alpha_\sigma=\pi/12$:
\begin{eqnarray}
w^{(+)}_{\sigma}&=&
\frac{\left(e^{\Omega_{\sigma}t -i\pi/12}+ 
e^{-\Omega_{\sigma}t +i\pi/12} \right)}{\sqrt{2\Omega_{\sigma}} 
N({\bf x})}F_{\sigma}({\bf x}),
\;\;\; \Omega_{\sigma} > 0, \nonumber \\
\label{intmodes2b}
\end{eqnarray}
in order to make $w^{(+)}_{\sigma}$ look ``as similar as possible" 
to $v^{(+)}_{\sigma}$. Because $w_\sigma^{(+)}$ and $w_\sigma^{(-)}$ 
grow exponentially in time, we borrow from cosmology the ``tachyonic" 
term (see, e.g., Ref.~\cite{FKL01}) and refer to these modes 
accordingly. (It should be noted, however, that in the cosmological context 
the scalar field is {\em self-interacting}, as ruled by some 
interacting potential, in contrast to our present case where 
it is {\em free}.)

As a result, the field operator $\hat \Phi (x)$ can be constructed 
using $\{v_\sigma^{(+)}, v_\sigma^{(-)} \}$ and 
$\{w_\sigma^{(+)}, w_\sigma^{(-)} \}$ as
\begin{eqnarray}
\hat{\Phi}  
&=& 
\int d\mu(\sigma)
[\hat b_{\sigma} v^{(+)}_{\sigma} + \hat b^{\dagger}_{\sigma} v^{(-)}_{\sigma}] 
\nonumber \\
&+& 
\sum_{\sigma} [\hat c_{\sigma} w^{(+)}_{\sigma} + \hat c_{\sigma}^{\dagger} w^{(-)}_{\sigma} ], 
\label{Phi_expandido2}
\end{eqnarray}
where $[\hat b_\alpha, \hat b_\beta^\dagger] = \delta (\alpha, \beta)$,
$[\hat c_\alpha, \hat c_\beta^\dagger] = \delta (\alpha, \beta)$
(with the other commutators vanishing),
and we have used the summation symbol in the right-hand side of 
Eq.~(\ref{Phi_expandido2}) because the tachyonic modes will be labeled later
with quantum numbers assuming discrete values. We recall that the
vacuum state $| 0 \rangle$ satisfies 
$\hat b_\sigma |0\rangle = \hat c_\sigma |0\rangle = 0$ for all $\sigma$.  
We note that in contrast to the $v_\sigma^{(+)}$ and $v_\sigma^{(-)}$
modes, $w_\sigma^{(+)}$ and $w_\sigma^{(-)}$ are not 
frequency eigenstates of $i \partial/\partial t$. 
As a result, the vacuum $|0\rangle$ and the other Fock states 
do not have in general any natural particle-content interpretation
(see Refs.~\cite{SS70} and~\cite{L12} for a field-theoretic 
discussion on the Fock space in the presence of tachyonic modes). 
This feature will lead us to use Unruh-DeWitt detectors to probe
vacuum fluctuations of the scalar field in Sec.~\ref{detectors}. 

Nevertheless,  important pieces of information are directly provided
through the (formal) stress-energy tensor operator: 
\begin{eqnarray}
\hat T_{a b}
&=&
  (1-2\xi)\nabla_a \hat \Phi  \nabla_b \hat \Phi 
+ \xi R_{ab} \hat \Phi^2 
- 2\xi \hat\Phi \nabla_a\nabla_b \hat\Phi 
\nonumber \\
&+& (2 \xi -1/2) g_{a b} 
[\nabla_c \hat\Phi \nabla^c \hat\Phi + (m^2 + \xi R){\hat\Phi}^2]
\end{eqnarray}
and the  corresponding Hamiltonian:
\begin{equation}
\hat{H} 
\equiv
\int_{\Sigma_t} d \Sigma_a \;\kappa_b \; \hat{T}^{ab}
\label{general_hamiltonian} 
 =  
\int_{\Sigma_t} d \Sigma \; \hat{\rho},
\end{equation}
where 
$d\Sigma_a \equiv d\Sigma \, n_a$, $\kappa_b = (\partial_t)_b$,
\begin{equation}
\hat \rho \equiv n_a \kappa_b \hat{T}^{a b}
\label{rho}
\end{equation}
is the energy-density operator in $\Sigma_t$ associated with the timelike isometry, 
and we recall that Eq.~(\ref{general_hamiltonian}) does not depend on 
the $\Sigma_t$ choice because $\nabla_a ( \kappa_b \hat T^{a b}) =0$. 
Thus, the total energy is conserved. By using 
Eq.~(\ref{Phi_expandido2}) in the Hamiltonian operator~(\ref{general_hamiltonian}), 
we obtain
\begin{eqnarray}
\hat{H}
&\equiv &  
\int_{\Sigma_t} d \Sigma\, N^{-1}\, \hat{T}_{00}
\nonumber \\
&=&
\frac{1}{2} \int d\mu(\sigma) 
(\hat b^{\dagger}_{\sigma} \hat b_{\sigma}
 + \hat b_{\sigma} \hat b^{\dagger}_{\sigma})\varpi_{\sigma}
+\sum_{\sigma} \hat{\cal H}_\sigma,
\label{hamiltonian_operator}
\end{eqnarray}
where
\begin{equation}
\hat{\cal H}_\sigma 
\equiv
-[{\sqrt{3}}/{2} (\hat c^{\dagger}_{\sigma} \hat c_{\sigma} 
+ \hat c_{\sigma}  \hat c^{\dagger}_{\sigma} ) 
+ \hat c_{\sigma} \hat c_{\sigma} + 
\hat c^{\dagger}_{\sigma}  \hat c^{\dagger}_{\sigma} ]\Omega_{\sigma}.
\label{H_sigma}
\end{equation}
In  contrast to the first term in the right-hand side of Eq.~(\ref{hamiltonian_operator}), 
associated with the oscillatory $v_\sigma^{(\pm)}$ modes, which always gives 
a positive-definite contribution to the  energy expectation value for {\em every} 
state choice, the second term, associated with the tachyonic modes $w_\sigma^{(\pm)}$, 
gives a negative contribution to the energy expectation
value for {\em some} states. This can be easily seen by rewriting Eq.~(\ref{H_sigma}) 
as
\begin{equation}
\hat{{\cal H}}_{\sigma} 
 = 
 (1- \sqrt{3}/2 ) \hat p_{\sigma}^2 - (1+ \sqrt{3}/2) \Omega_{\sigma}^2 \hat q_{\sigma}^2,
\label{harmonic_oscillator_up_side_down}
\end{equation}
where we have defined the position- and momentum-like operators 
\begin{eqnarray}
\hat{q}_{\sigma} 
&\equiv & \frac{1}{\sqrt{2\Omega_{\sigma}}}
(\hat c_{\sigma} + \hat c^{\dagger}_{\sigma} ), 
\label{qn} \\
\hat{p}_{\sigma} 
&\equiv & 
i\sqrt{\frac{\Omega_{\sigma}}{2}} (\hat c^{\dagger}_{\sigma} - \hat c_{\sigma} ),
\label{pn}
\end{eqnarray}
respectively, satisfying $[\hat{q}_{\sigma},\hat{p}_{\sigma} ]=i \, \hat{I}$
with $\hat{I}$ being the identity operator. 
Equation~(\ref{harmonic_oscillator_up_side_down}) is formally identical to 
the Hamiltonian of a non-relativistic particle in a harmonic potential 
turned upside down~\cite{B86}. It is clear, then, that the ``potential" 
term gives a negative contribution to the energy expectation value. 
In particular, for states  $|\Psi  \rangle $ satisfying
\begin{equation}
\hat c_\sigma^\dagger \hat c_\sigma |\Psi \rangle = \Xi |\Psi \rangle,
\,\,\, \Xi \in \mathbb{N},
\label{Psi_tachyonic}
\end{equation}
which include the vacuum state, it is easy to see that
$
\langle \Psi| \hat c_\sigma \hat c_\sigma + 
\hat c_\sigma^\dagger \hat c_\sigma^\dagger|\Psi \rangle 
=0
$
and, thus, 
\begin{equation}
\langle \Psi|\hat{\cal H}_\sigma |\Psi\rangle 
= - \sqrt{3}( 1/2 + \Xi ) \, \Omega_\sigma < 0.
\label{unbounded_from_below}
\end{equation}
Hence, for these states the negative contribution coming from 
the ``potential" in Eq.~(\ref{harmonic_oscillator_up_side_down}) 
dominates over the corresponding positive one coming from the 
``kinetic" term. (We shall return to this point when we discuss the 
excitation of Unruh-DeWitt detectors in Sec.~\ref{detectors}.) 
The fact that the right-hand side of Eq.~(\ref{unbounded_from_below})
may be arbitrarily negative for sufficiently large $\Xi $ 
reflects the fact that $\hat{\cal H}_\sigma$ is unbounded
from below.


\section{Awaking the vacuum in relativistic stars due to tachyonic instability}
\label{awaking_the_vacuum_in_stars}

Now, we shall see how tachyonic modes can appear in neutron-like stars and 
discuss their consequences~\cite{LV10,LMV10}. Let us assume the case in which 
(A)~classical matter initially scattered throughout space with very low 
density eventually collapses to form (B)~a static and stable 
star according to general relativity. Spacetimes associated with 
situations~A-B are well described by the line elements 
[see Eq.~(\ref{static line element})]
\begin{eqnarray}
ds^2=  \left\{
 \begin{array}{ll}
                    -dt^2 + d {\bf x}^{2}   & \,\,\, {\rm (A)} \\
      N^2_{\rm (B)}(-dt^2 + h^{\rm (B)}_{i j} dx^i dx^j) &\,\,\, {\rm (B)} \\
  \end{array} 
        \right..
 \label{metric0}      
 \end{eqnarray}
We note that for the time being we will restrict our investigation to the static 
regions~A and~B of the spacetime. Comments about how our present analysis can be 
completed as one takes into account the time evolution between the static 
eras are made along the text. For the sake of obtaining explicit 
results, we make an extra simplification in this section and 
consider spherically symmetric stars  in which case Eq.~(\ref{metric0}) is 
replaced by 
\begin{equation}\label{star_formation_metric}
 ds^2 = \left\{\begin{array}{ll}
-dt^2 + d{\bf x}^2 &\textrm{(A)}\\
-f(dt^2-d\chi^2)+r^2(d\theta^2+\sin^2\theta\, d\varphi^2) &\textrm{(B)}
\end{array}
\right.,
\end{equation}
where $f=f(\chi)>0$ and $r=r(\chi)\geq 0$ satisfy $f(\chi)\to 1$  and  
$r(\chi)/\chi \to 1$ for $\chi\to \infty$, and $dr/d\chi>0$ so that no 
trapped light-like surface is present.

We construct the field operator similarly as in Eq.~(\ref{Phi_expandido}):
\begin{equation}
\hat \Phi 
= 
\int d^3 k
[\hat a_{\bf k} u_{\bf k}^{(+)} +\hat a_{\bf k}^\dagger u_{\bf k}^{(-)}],
\label{Phi_expandido_asymptotic_past}
\end{equation} 
where we choose here $u^{(\pm)}_{\bf k}$ such that they assume the usual 
flat-space stationary form in the asymptotic past (region~A):
\begin{equation}\label{u_modes_star}
u^{(\pm)}_{\bf k}\stackrel{{\rm (A)}}{=}
(16\pi^3\omega_{\bf k})^{-1/2} \exp[\mp i(\omega_{\bf k}t - {\bf k}\cdot{\bf x})]
\end{equation}
with ${\bf k}\in\mathbb{R}^3$ and $\omega_{\bf k}\equiv \sqrt{{\bf k}^2+m^2}$. 
This choice is motivated by the fact that we shall assume hereafter the scalar 
field to be in the no-particle state   $|0\rangle_{\rm in}$ as described by 
static observers in the asymptotic past: $a_{\bf k} |0\rangle_{\rm in}=0$.

Now, let us represent $\hat \Phi$ in terms of positive- and negative-norm
modes in region~B, when the star has settled down, as 
[see Eq.~(\ref{Phi_expandido2})]
\begin{eqnarray}
\hat{\Phi}  
&=& 
\sum_{ l \mu} \int d\varpi 
[\hat b_{\varpi \,l \mu} v^{(+)}_{\varpi \,l \mu} 
+ \hat b^{\dagger}_{\varpi \,l \mu} v^{(-)}_{\varpi \,l \mu}] 
\nonumber \\
&+& 
\sum_{\Omega \,l \mu} 
[\hat c_{\Omega \, l \mu} w^{(+)}_{\Omega \,l \mu} 
+ \hat c_{\Omega \,l \mu}^{\dagger} w^{(-)}_{\Omega l \mu} ], 
\label{Phi_expandido2_estrela}
\end{eqnarray}
where~\cite{errata}
\begin{equation}
v^{(+)}_{\varpi l \mu} 
\stackrel{{\rm (B)}}{=}
e^{-i \varpi t} 
\frac{F_{\varpi l}(\chi)}{\sqrt{2\varpi\,}  r(\chi)}  
Y_{l \mu} (\theta, \phi), 
\;\;\; \varpi > 0,
\label{intmodes_estrela}
\end{equation}
\begin{eqnarray}
w^{(+)}_{\Omega l \mu} 
& \stackrel{{\rm (B)}}{=} &
( e^{ \Omega t -i\pi/12} + e^{-\Omega t +i\pi/12} )
\nonumber \\
&\times& 
\frac{F_{\Omega l}(\chi)}{\sqrt{2\Omega\,} r(\chi)} Y_{l \mu} (\theta, \phi),
\;\;\; \Omega > 0.
\label{intmodes2_estrela}
\end{eqnarray}
Here, $Y_{l \mu} (\theta, \phi)$ are the usual spherical harmonics 
($l=0,1,2,\dots$ and $\mu=-l,-l+1,\dots,l$),  
$F_{\varpi l}(\chi)$ and $F_{\Omega l} (\chi)$ satisfy 
\begin{equation}\label{schrodinger_eq}
-\frac{d^2}{d\chi^2} F_{\varpi l} +V_{\rm eff}^{(l)} F_{\varpi l}= 
\varpi^2 F_{\varpi l}
\end{equation}
and
\begin{equation} \label{schrodinger_eq2}
-\frac{d^2}{d\chi^2} F_{\Omega l} +V_{\rm eff}^{(l)} F_{\Omega l}= 
- \Omega^2 F_{\Omega l},
\end{equation}
respectively, and 
\begin{equation}\label{eff_potential_star}
V_{{\rm eff}}^{(l)}\equiv f\left( m^2+\xi
R+\frac{l(l+1)}{r^2}\right)+\frac{1}{r}\frac{d^2r}{d\chi^2}
\end{equation}
is the effective potential. For perfect-fluid stars, the effective 
potential~(\ref{eff_potential_star}) can be cast as
\begin{equation}\label{eff_potential_density}
V_{\rm eff}^{(l)}
=
f 
\left[ 
m^2+\frac{l(l+1)}{r^2} + \left(\xi-\frac{1}{6} \right)R
+\frac{8\pi G}{3}(\bar{\rho}-\rho)
\right],
\end{equation}
where $\rho = \rho (\chi )$ is the mass-energy density of the 
stellar fluid and
\begin{equation}\label{energy_density_average}
\bar{\rho}(\chi)
\equiv \frac{3M(\chi)}{4\pi [r(\chi)]^3} 
\end{equation} 
is the corresponding average density up to the radius coordinate $r(\chi)$, 
which encompasses a mass $M(\chi)$. We remind that according 
to general relativity $R=8\pi G(\rho-3p)$, where $p$ is the hydrostatic 
pressure which bears the star up against its weight.

As discussed in Sec.~\ref{standard_quantization_static_spacetimes}, 
the appearance of tachyonic modes in the present context will depend 
on the existence of nontrivial solutions for Eq.~(\ref{schrodinger_eq2}). 
They are expected to exist for negative enough effective potentials 
satisfying  
\begin{equation}
|V_{\rm eff}^{(l)}| R_s^2 \gtrsim 1, 
\label{VL2}
\end{equation}
where $r= R_s$ is the star radius. Because the centrifugal barrier 
in Eq.~(\ref{eff_potential_density}) is positive, we look for tachyonic
solutions of Eq.~(\ref{schrodinger_eq2}) with $l=0$ which are the most 
likely ones to exist (if any). By taking $f \sim 1$ and assuming 
$\bar \rho \sim \rho$, we obtain from Eq.~(\ref{eff_potential_density}) that 
$$
V_{\rm eff}^{(0)} \sim m^2 +  (\xi-1/6) R.
$$ 
Clearly, only the second term in the right-hand side can be negative. Then, 
Eq.~(\ref{VL2}) implies that a necessary condition for the
existence of tachyonic modes with $\xi \approx 1$ is 
\begin{equation}\label{density_radius_condition}
\frac{\rho}{10^{15}\ {\rm g}/{\rm cm}^3}
\left( \frac{R_s}{7\ {\rm km}} \right)^2 \gtrsim 1
\end{equation}
and
\begin{equation}\label{field_mass_condition}
\frac{m^2/(3.5\times 10^{-11}\ {\rm eV})^2}{\rho/(10^{15}\
{\rm g}/{\rm cm}^3)}\ll 1,
\end{equation}
where we have set $\rho \sim p$. 
Equations~(\ref{density_radius_condition}) and~(\ref{field_mass_condition}) show 
that the appearance of tachyonic modes for small $\xi$ values in the spacetime 
of typical neutron stars requires the scalar field to be light: 
$m \lesssim 10^{-11}\, {\rm eV}$. 

Although light scalars are widely considered in astrophysics and cosmology, there 
is the issue about how much extra mass they could acquire from Planck-scale 
radiative corrections. For axions, e.g., a general expression for
the mass shift can be cast as $ \delta m_a^2 \sim K_a f_a^{n+2}/M^n_P$, 
where $K_a$ is some unknown coupling constant, $f_a \sim 10^{12} \, {\rm GeV}$ is 
the energy scale of the Peccei-Quinn symmetry breaking, $M_P \sim 10^{19} \, 
{\rm GeV}$ is the Planck energy, and $n$ is a model-dependent positive integer 
(associated with the dimension of the symmetry-breaking operators appearing 
in the effective Lagrangian)~\cite{P00}. We see, then, that $\delta m_a$ can 
easily exceed, e.g., $10^{-5}\, {\rm eV}$  (ruling out axions as a dark matter 
candidate) unless $K_a$ and $n$ turn out to be small and large enough, 
respectively (see Ref.~\cite{axion1} and references therein). 
Fortunately, explicit models showing how scalar fields can be protected 
from acquiring large mass due to quantum gravity effects have already 
been worked out (see, e.g., Ref.~\cite{axion2}). 
In our context, assuming the electroweak symmetry breaking of the standard 
model which has an energy scale of $\Lambda_{\rm ESM} \sim 100 \,{\rm GeV}$, 
the corresponding mass shift would be $ \delta m^2 \sim K 
\Lambda_{\rm ESM}^{n+2} /M^n_P$, where again $K$ and $n$ are unknown. 
Here, we pragmatically {\em assume} that Planck-scale effects will not shift 
the mass of our scalar field beyond $10^{-12}\, {\rm eV}$. At this point, it 
is difficult to say how strong this assumption is because of our lack of 
understanding of the Planck-scale physics. Still, this is much less demanding 
than what is usually required for quintessence fields where the mass shift 
cannot typically exceed the mass scale defined by the Hubble constant 
$ 10^{-33} \, {\rm eV}$~\cite{FHSW95}. A detailed analysis of this issue 
would be welcome but goes far beyond the scope of this 
paper. For computational purposes, we take our scalar field to be massless.

By assuming stars with uniform and parabolic density profiles
and suitable $\xi$ values (typically $\xi>1/6$ and $\xi\lesssim-2$),
it was shown in Ref.~\cite{LMV10} that tachyonic modes do appear 
for  $M/R_s$ ratios compatible with neutron-like stars.

Now, let us proceed by recalling that the positive-norm in-modes 
$u_{\bf k}^{(+)}$, which in region~A look like as exhibited in 
Eq.~(\ref{u_modes_star}), will emerge, in general, as a combination 
of positive- and negative-norm modes  
$\{v^{(+)}_{\varpi l \mu}, v^{(-)}_{\varpi l \mu}\}$ and
$\{w^{(+)}_{\Omega l \mu},w^{(-)}_{\Omega l \mu}\}$ in region~B
[see Eqs.~(\ref{intmodes_estrela}) and~(\ref{intmodes2_estrela})].
Hence, not only the in-vacuum will not coincide in general with the out-vacuum 
but also at least some of the in-modes will certainly go through a phase of 
exponential  growth provided, of course, the existence of 
tachyonic modes $w^{(\pm)}_{\Omega l \mu}$. This leads to what was denominated 
{\em vacuum awakening effect in relativistic stars}, i.e., an exponential 
amplification of the vacuum fluctuations~\cite{LV10,LMV10}. In 
order to see this, we use Eq.~(\ref{Phi_expandido2_estrela}) to calculate  
\begin{equation}\label{quantum_fluctuations}
_{\rm in}\langle0|\hat{\Phi}^2|0\rangle_{\rm in}
\stackrel{{\rm (B)}}{\sim}
\kappa\frac{e^{2\bar{\Omega\,} t}}{8\pi \bar{\,\Omega}}
\left(\frac{{F}_{\bar \Omega 0} (\chi)}{r(\chi)}\right)^2
[1+ {\cal O}(e^{-\epsilon t})],
\end{equation}
where ${F}_{\bar{\Omega} 0} (\chi)$ denotes the solution of 
Eq.~(\ref{schrodinger_eq2}) with the most negative eigenvalue, 
$-\bar{\Omega}^2$ (taking $l=0$, which is the most favorable case), 
$\epsilon = {\rm const}>0$, 
and $\kappa$ is a constant of order unity whose value depends 
on (i)~projections of modes $u^{(\pm)}_{\bf k}$ on $w_{\bar \Omega l \mu}^{(\pm)}$
and (ii)~the quantum state, assumed here to be the in-vacuum 
$|0\rangle_{\rm in}$. It is worthwhile to emphasize that ultraviolet 
divergences, which should be renormalized to obtain $\langle \hat \Phi^2 \rangle$, 
are associated with the $\varpi \to \infty$ sector of the oscillatory modes [see 
Eq.~(\ref{Phi_expandido2_estrela})] and does not concern the tachyonic 
modes ($\Omega^2 \leq \bar \Omega^2 < \infty$), which are the ones 
giving the dominant contribution in Eq.~(\ref{quantum_fluctuations}) (because of the 
$\exp (2 \bar \Omega t)$ term). Accordingly, the expectation value of 
the vacuum energy density~(\ref{rho}), namely,
$$
 _{\rm in}\langle 0| \hat{\rho} |0\rangle_{\rm in} 
 \equiv n^a \kappa^b \,_{\rm in}\langle 0|\hat{T}_{ab}|0\rangle_{\rm in},
$$ 
experiences an exponential amplification:
\begin{eqnarray}\label{energy_density}
_{\rm in}\langle 0| \hat{\rho} |0\rangle_{\rm in} 
&&
\stackrel{{\rm (B)}}{\sim}
\kappa \frac{\bar{\Omega} e^{2\bar{\Omega} t}}{16\pi\sqrt{f}}  
\left\{ \frac{1-4\xi}{2r^2} \frac{d}{d\chi} 
\left(r^2\frac{d}{d\chi} \left(\frac{{F}_{\bar \Omega 0}}{\bar{\Omega}r}\right)^2\right) \right.\nonumber\\
&&+\frac{\xi}{\bar{\Omega}^2r^2}
\frac{d}{d\chi}\left(\frac{{{F}_{\bar \Omega 0}}^2}{f}\frac{df}{d\chi}\right)\Bigg\}
[1+ {\cal O}(e^{-\epsilon t})].
\end{eqnarray}
The time scale which rules how fast the vacuum energy density increases
is given by $\bar \Omega^{-1} \sim |V_{\rm eff}^{(0)}|^{-1/2} \sim R_s$ 
[see Eq.~(\ref{VL2})]. By using this, we rewrite Eq.~(\ref{energy_density}) as 
\begin{eqnarray}
_{\rm in}\langle 0| \hat{\rho} |0\rangle_{\rm in} 
& \stackrel{{\rm (B)}}{\sim} &
\bar \Omega h(\bar r) e^{2 \bar \Omega t}/ R_s^3
\nonumber \\
& \stackrel{{\rm (B)}}{\sim}& 
h(\bar r) 
\exp \left[ \frac{t/(10^{-5} s)}{R_s/ (10 {\rm km})} \right] 
\frac{10^{-62} {\rm g/cm}^3}{R_s^4/(10 {\rm km})^4},
\nonumber \\
\label{estimativa}
\end{eqnarray}
where $h(\bar r)$ is a dimensionless function of $\bar r \equiv r/R_s$
(which vanishes asymptotically and is of order unity for $\bar r \sim 1$). 
We see from Eq.~(\ref{estimativa}) that once 
the effect is triggered by a neutron-like star with $R_s \approx 10~{\rm km}$, 
few milliseconds would be enough for the vacuum energy density to 
become dominant over the star classical mass-energy density (which can 
be as high as $10^{14} - 10^{17} {\rm g/cm}^3$). We must emphasize that 
at some point the spacetime must backreact against the growth of the vacuum energy 
density, affecting the field  and ceasing the instability by taming 
the tachyonic modes. Eventually, field and spacetime must reach a new stable 
configuration.  As argued in Ref.~\cite{PCBRS11}, one possibility would
be that for some values of $\xi$ the spontaneous scalarization 
mechanism~\cite{DE93} could restabilize the star. In our context, this would 
correspond to a symmetry breaking which would lead $ \langle \hat{\Phi}\rangle$, 
which is null as calculated in the $|0\rangle_{\rm in}$ vacuum state, to 
acquire a nonzero large value  compatible with the exponentially 
amplified $ \langle \hat{\Phi}^2 \rangle$ [see Eq.~(\ref{quantum_fluctuations})].
Whether the star will end up destroyed or somehow rebalanced is 
unknown at this point.

We close this section explaining how the vacuum energy density 
amplification is consistent with energy conservation discussed 
below Eq.~(\ref{rho}). For this purpose, let us
note that $\nabla_a (\kappa_b \hat T^{a b} )=0$ can be rewritten
as
\begin{equation}
\partial_t \hat \rho + \frac{1}{\sqrt{^{(3)}g\,}} \partial_i (\sqrt{^{(3)}g\, } \,{\hat{j}\,}^i)=0,
\label{conservaccao}
\end{equation}
where ${\hat{j}\,}^i \equiv -\sqrt{f} \, \hat T^i_{\, 0}$ is the energy-current 
density. The corresponding vacuum expectation value can be calculated and reads
\begin{equation}\label{energy_current}
_{\rm in}\langle 0|{\hat{j}\,}^i|0\rangle_{\rm in}
\stackrel{{\rm (B)}}{\sim} 
-\bar \Omega  V^i (\bar r) e^{2\bar \Omega t}/R_s^3,
\end{equation} 
where 
$
^{(3)} \nabla_i V^i 
\equiv ( {^{(3)}g}\,)^{-1/2} \partial_i (\sqrt{^{(3)}g\, } \,V^i)
= 2 \bar \Omega h(\bar r)
$.
Thus, the total energy is conserved because the gravitational field
redistributes the vacuum energy density in such a way that  
an amplification of $_{\rm in}\langle0| \hat \rho |0 \rangle_{\rm in}$ 
somewhere with positive magnitude must be compensated elsewhere by a 
corresponding amplification with negative magnitude.

\section{Probing the awoken phase using detectors}
\label{detectors}

Now, in order to probe the building up of the vacuum energy density
in region~B, where the vacuum is awake by the presence of tachyonic modes, 
we will look directly at the response of Unruh-DeWitt detectors. We shall do so
because, as discussed in Sec.~\ref{standard_quantization_static_spacetimes},
the Fock-space states have no natural particle-content interpretation
[see discussion  below Eq.~(\ref{Phi_expandido2})]. Here, we relax the 
spherical symmetry assumption of the previous section and consider 
regions~A and~B as described by the line elements~(\ref{metric0}). Because 
we want to avoid any contributions in the response coming from the motion 
of the apparatus, the detector (with proper time $\tau$) is made to lie 
static following an integral curve $x=x(\tau)$ of the timelike isometry. 

We consider here a two-level Unruh-DeWitt detector represented 
by a Hermitian operator $\hat{m}_0$ acting in a Hilbert space 
spanned by unexcited and excited energy eigenstates 
$|E_0\rangle$ and $|E\rangle$ ($E>E_0$), respectively. The 
detector is prepared to be initially unexcited. For our 
purposes, it is convenient to switch it on in the very 
beginning of region~B, where we set $\tau \equiv 0$. 
At the tree level, the excitation probability as a function 
of the proper-time interval $T$ is given by~\cite{BD82}  
\begin{equation}
P_{\rm exc}=|\langle E|\hat{m}_0|E_0\rangle|^2\mathcal{F}(\Delta E),
\end{equation}
where $\Delta E \equiv E - E_0$ and the response 
function is  
\begin{equation}
\mathcal{F}(\Delta E)=\int_{0}^{T}  d\tau \int_{0}^{T}  d\tau' 
e^{-i\Delta E (\tau-\tau')}
G^+_{\rm in}[x(\tau), x(\tau')]
\label{response}
\end{equation}
with
$$
G^+_{\rm in} [x(\tau), x(\tau')] 
\equiv 
{_{\rm in}\langle} 0|\hat{\Phi}[x(\tau)] \hat{\Phi} [x(\tau')] |0 \rangle_{\rm in}.
$$ 
The two-point  function is written throughout the spacetime in terms of 
the in-modes as
\begin{equation}
G^+_{\rm in}[x,x'] = \int d^3k\; u^{(+)}_{{\bf k}}(x)\, u^{(-)}_{{\bf k}}(x'),
\label{twopoint}
\end{equation}
where we recall that  in the asymptotic past $u^{(\pm)}_{{\bf k}}(x)$ take the 
simple form given in Eq.~(\ref{u_modes_star}). Next, we suitably 
decompose $u^{(+)}_{\bf k}$ in terms of~$\{v^{(\pm)}_\sigma \}$ 
and~$\{ w^{(\pm)}_\Omega \}$ as
\begin{eqnarray}
u^{\rm (+)}_{{\bf k}} 
& = &
\alpha^*_{\Omega {\bf k}}w^{(+)}_{\Omega} 
- \beta_{\Omega {\bf k}}w_{\Omega}^{(-)}
\nonumber \\
& + &
\int d\mu(\sigma)(\alpha^*_{\sigma {\bf k}} v_{\sigma}^{(+)} 
- \beta_{\sigma {\bf k}}v_{\sigma}^{(-)}),
\label{bog}
\end{eqnarray} 
where, for the sake of simplicity, we have assumed that the scalar field 
is made unstable in region~B by the existence 
of a single tachyonic mode [see Eq.~(\ref{intmodes2b})]:
\begin{equation}
w^{(+)}_{\Omega} \stackrel{\rm (B)}{=}
\frac{\left(e^{\Omega t -i\pi/12}+ 
e^{-\Omega t +i\pi/12} \right)}{\sqrt{2\Omega \,} 
N_{\rm (B)}({\bf x})}F_\Omega^{\rm (B)} ({\bf x}),
\;\; \Omega> 0. 
\label{intmodes2}
\end{equation}
The tachyonic mode above will be labeled by $\Omega$ (with the other 
quantum numbers being omitted to simplify notation), while the oscillatory
modes are cast here as [see Eq.~(\ref{intmodes0})]
\begin{equation}
v^{(+)}_{\sigma} \stackrel{{\rm (B)}}{=} 
\frac{e^{-i \varpi_{\sigma}t}}{\sqrt{2\varpi_{\sigma} \,} 
N_{\rm (B)}({\bf x})}
F^{\rm (B)}_{\sigma}({\bf x}), 
\;\; \varpi_{\sigma} > 0.
\label{intmodes}
\end{equation}
The Bogoliubov coefficients in Eq.~(\ref{bog}) are
\begin{eqnarray}
&&
\alpha_{\sigma {\bf k}} = (u^{(+)}_{{\bf k}},v^{(+)}_\sigma )_{\rm KG}, 
\,\,\,
\beta_{\sigma {\bf k}} = -(u^{(-)}_{{\bf k}},v^{(+)}_\sigma )_{\rm KG}, 
\nonumber \\ 
&&
\alpha_{\Omega\, {\bf k}} = (u^{(+)}_{{\bf k}},w^{(+)}_{\Omega} )_{\rm KG}, 
\,\,\,
\beta_{\Omega\, {\bf k}} = -(u^{(-)}_{{\bf k}},w^{(+)}_{\Omega} )_{\rm KG}
\nonumber
\end{eqnarray}
as calculated in {\em any} Cauchy surface.

Now, we  must proceed and evaluate  the response function~(\ref{response}). 
For this purpose, it is convenient to use
Eqs.~(\ref{bog})-(\ref{intmodes}) to calculate
\begin{eqnarray}
&& \int_{0}^{T} d\tau e^{-i\Delta E \tau} u^{(+)}_{\bf k}[x(\tau)]  = \int  d\mu(\sigma)
\nonumber \\
&& 
\times
\left[
\alpha^*_{\sigma {\bf k}} 
\frac{F^{\rm (B)}_{\sigma}}{N_{\rm (B)}} 
\psi_{\sigma +}  
-  
\beta_{\sigma {\bf k}} 
\frac{F^{{\rm (B)}*}_{\sigma}}{N_{\rm (B)}} 
\psi^*_{\sigma -} 
\right]_{{\bf x}= {\bf x}_d} 
\nonumber \\
&& 
+ 
\left[ 
\alpha^*_{\Omega {\bf k}}
\frac{F_\Omega^{\rm (B)}}{N_{\rm (B)}}
\Psi_{+} 
- \beta_{\Omega {\bf k}} 
\frac{F_\Omega^{{\rm (B)}*}}{N_{\rm (B)}}
\Psi^*_{-}
\right]_{{\bf x}= {\bf x}_d}, 
\label{intu}
\end{eqnarray}
where we have defined 
\begin{eqnarray}
\Psi_\pm \!\!\!
&\equiv&
\int_0^T  d\tau 
\frac{e^{\mp i\Delta E \tau} 
      \left(
       e^{\Omega\tau/N_{\rm (B)}-i\pi/12}
      +e^{-\Omega\tau/N_{\rm (B)} +i\pi/12}
      \right)}{\sqrt{2 \Omega}}
\nonumber \\
&=&
\frac{1}{\sqrt{2 \Omega}}
\biggl[\frac{e^{i \pi/12} (1-e^{-(\Omega/N_{\rm (B)}\pm i \Delta E)T})}{\Omega/N_{\rm (B)}\pm i \Delta E}
 \nonumber \\
&&
+ \frac{e^{-i \pi/12}(e^{(\Omega/N_{\rm (B)}\mp i \Delta E)T}-1)}{\Omega/N_{\rm (B)}\mp i \Delta E}\biggr],
\label{Psi}
\end{eqnarray}
\begin{eqnarray}
 \psi_{\sigma \pm}
&\equiv& 
\frac{1}{\sqrt{2 \varpi_\sigma}} \int_0^Td \tau 
    \; e^{-i\left(\varpi_\sigma /N_{\rm (B)} \pm \Delta E\right)\tau}
\nonumber \\
&=&
\frac{2 e^{-i ( \varpi_{\sigma}/N_{\rm{(B)}}\pm \Delta E ) T/2}}{\sqrt{2 \varpi_{\sigma}}}
\nonumber \\ 
&\times& 
\frac{\sin\left[(\varpi_{\sigma}/N_{\rm{(B)}}\pm \Delta E)T/2\right]}
{\varpi_{\sigma}/N_{\rm{(B)}}\pm \Delta E},
\label{psi}
\end{eqnarray}
and ${\bf x}= {\bf x}_d$ is the detector's spatial position. 
Then, we write the detector response function~(\ref{response})  
with the help of Eq.~(\ref{intu}) as
\begin{equation}
\mathcal{F}(\Delta E)= [\mathcal{F}_0 + \mathcal{F}_1 + \mathcal{F}_2]_{{\bf x}= {\bf x}_d},
\label{asymresp}
\end{equation}
where
\begin{equation}
\mathcal{F}_0= 
\int d^3 k 
\left|
\alpha^*_{\Omega {\bf k}} 
\frac{F_\Omega^{\rm (B)}}{N_{\rm (B)}}
\Psi_+  
- \beta_{\Omega {\bf k}} 
\frac{F_\Omega^{{\rm (B)}*}}{N_{\rm (B)}}
\Psi_-^*  
\right|^2,
\label{resp0}
\end{equation}
\begin{eqnarray}
 \mathcal{F}_1 & = & 
\int d^3 k 
\int d\mu(\sigma)
\nonumber \\
&& 2 {\rm Re} 
\left[ 
 \left( \alpha^*_{\Omega{\bf k}} 
 \frac{F_\Omega^{\rm (B)}}{N_{\rm (B)}}
 \Psi_+  
       -\beta_{\Omega{\bf k}}
      \frac{F_\Omega^{{\rm (B)}*}}{N_{\rm (B)}} 
       \Psi_-^*  
 \right)
\right.
\nonumber \\
&& \times 
 \left.
 \left( \alpha_{\sigma {\bf k}}
 \frac{F_\sigma^{{\rm (B)}*}}{N_{\rm (B)}}
 \psi^*_{\sigma +}
       -\beta^*_{\sigma {\bf k}} 
       \frac{F_\sigma^{\rm (B)}}{N_{\rm (B)}}
       \psi_{\sigma -} 
 \right)  
\right],\nonumber \\
\end{eqnarray}
\begin{equation}
\mathcal{F}_2= 
\int d^3 k 
\left| 
\int d\mu(\sigma)
\left(
\alpha^*_{\sigma {\bf k}} 
\frac{F_\sigma^{\rm (B)}}{N_{\rm (B)}}
\psi_{\sigma +}
-\beta_{\sigma {\bf k}} 
\frac{F_\sigma^{{\rm (B)} *}}{N_{\rm (B)}}
\psi^*_{\sigma -}
\right)
\right|^2.
\label{resp2} 
\end{equation}

The physical meaning of the response $\mathcal{F} (\Delta E)$ is 
more easily grasped in the case where the proper time interval $T$ is 
``large", i.e., $T \gg \Delta E^{-1}$ (in addition to 
$T \gg \Omega^{-1}$ whenever the tachyonic mode is present) 
and we will assume this hereafter up to the end of this section. 
In this case, Eqs.~(\ref{Psi}) and~(\ref{psi}) can be written as
\begin{eqnarray}
\Psi_{\pm}
\approx 
\frac{e^{-i(\pi/12 \pm \Delta E)T}e^{T \Omega/N_{\rm (B)}}}
{\sqrt{2\Omega}(\Omega/N_{\rm (B)}\mp i \Delta E)}
\label{asymPsi}
\end{eqnarray}
and
\begin{eqnarray}
\psi_{\sigma \pm} 
& \approx &
\sqrt{\frac{2\pi^2}{\varpi_{\sigma}}}
e^{-i(\varpi_{\sigma}/N_{\rm{(B)}}
\pm \Delta E)T/2}
\nonumber \\
& \times & \delta ( {\varpi_{\sigma}}/{N_{\rm{(B)}}}\pm \Delta E),
\label{asympsi}
\end{eqnarray}
respectively.  
For the sake of comparison, we shall discuss separately the situations where 
the tachyonic mode $w^{(+)}_\Omega$ is present from the one where it is absent.

In the case where the tachyonic mode $w^{(+)}_\Omega$ is absent, 
$F_\Omega^{\rm (B)} =0 $ and, thus, the detector response becomes simply
$$
\mathcal{F} (\Delta E) = \mathcal{F}_2|_{{\bf x}= {\bf x}_d}.
$$
This is the usual result whose interpretation is straightforward: 
assuming that the detector stays switched 
on for an arbitrarily long time $T$,  we have from Eq.~(\ref{asympsi}) that 
$\psi_{\sigma +} \approx 0$ and according to Eq.~(\ref{resp2}) the detector 
excites by the absorption of particles created due to the  spacetime transition 
from regions~A to~B ($\beta_{\sigma {\bf k}} \neq 0$). (Note that when no 
restriction is posed on $T$, the detector excitation will also have a contribution 
coming from the process of switching it on and off  
($\psi_{\sigma +} \neq 0$)~\cite{HMP93}.) 

It is also interesting to note that in the absence of the tachyonic mode, 
the response will not grow faster than $T$. By recalling from Eq.~(\ref{asympsi}) 
that $\psi_{\sigma +} \approx 0$,  we write the response as [see Eq.~(\ref{resp2})]
\begin{eqnarray}
\mathcal{F}  (\Delta E) 
&
\! \approx \!
&
\int d^3 k 
\left| 
\int d\mu (\sigma )
\frac{F^{{\rm (B)}*}_{\sigma}}{N_{\rm (B)}}
\beta_{\sigma {\bf k}} \psi^*_{\sigma -}
\right|^2_{{\bf x}= {\bf x}_d}  
\nonumber \\
& 
\!\leq\!  
&
\int d^3 k 
\left(
\int d\mu (\sigma ) |\beta_{\sigma {\bf k}}|^2
\right)
\nonumber \\
& 
\!\times \!
&
\left(
\int d\mu (\sigma )
|\psi_{\sigma -}|^2 
\frac{{|{F_\sigma^{\rm (B)}}|^2 }}{N^2_{\rm (B)}}
\right)_{{\bf x}= {\bf x}_d},
\label{F2}
\end{eqnarray}
where we have used above the Cauchy-Schwarz inequality.
Now, by using Eq.~(\ref{psi}) [or directly Eq.~(\ref{asympsi})] in conjunction 
with the identity  
$\lim_{A\rightarrow \infty}\sin^2(\omega A)/\omega^2A= \pi \delta(\omega)$, 
we obtain
\begin{equation}
|\psi_{\sigma -}|^2
\approx 
(\pi T/\varpi_{\sigma})\, \delta(\varpi_{\sigma}/N_{\rm (B)}-\Delta E).
\label{aux}
\end{equation} 
This is used in Eq.~(\ref{F2}) to conclude indeed that
the response will not grow faster than $T$:
\begin{equation}
\mathcal{F} (\Delta E) \lesssim C_2 T
\label{FmenorT}
\end{equation}
with 
\begin{eqnarray} 
C_2 
& = &
\int d^3k 
\left(
\int d\mu (\sigma ) |\beta_{\sigma {\bf k}}|^2
\right)
\nonumber \\
& \times &
{\left(
\frac{\pi}{\Delta E} 
\int d\mu (\sigma ) \frac{|{F^{\rm (B)}_{\sigma}}|^2}{N^2_{\rm (B)}}
\delta ( \varpi_{\sigma}- N_{\rm (B)} \Delta E )
\right).}_{\!\!\!{\bf x}= {\bf x}_d}
\end{eqnarray}

On the other hand, assuming that the tachyonic mode 
$ w^{\rm (+)}_\Omega$ is present, the detector response 
will be dominated by Eq.~(\ref{resp0}) and, thus, 
$\mathcal{F} (\Delta E) \approx \mathcal{F}_0|_{{\bf x}= {\bf x}_d}$.
Thus,  we use Eq.~(\ref{asymPsi}) 
in Eq.~(\ref{resp0}) to obtain 
\begin{eqnarray}
&& 
\mathcal{F}(\Delta E)
\approx
Z  \, \exp [2 T \Omega/N_{\rm (B)} ({\bf x}_d)]
\nonumber \\
&& \times
\int d^3 k 
\left|
\alpha^*_{\Omega{\bf k}}
\frac{F_\Omega^{\rm (B)}}{N_{\rm (B)}}
e^{-i\pi/12} 
- 
\beta_{\Omega{\bf k}}
\frac{F_\Omega^{{\rm (B)}*}}{N_{\rm (B)}}
e^{i\pi/12}
\right|^2_{{\bf x}= {\bf x}_d}
\label{asymresp0simples} 
\end{eqnarray}
where  
$$
Z = \frac{1}
{2 \Omega [\Delta E^2 + \Omega^2/N^2_{\rm (B)}({\bf x}_d)]}.
$$
The exponential increase in the detector response reflects the growth
of the vacuum fluctuations and will continue as long as the unstable 
phase is not forced to terminate. This is possible because each excitation 
of the detector is accompanied by a corresponding decrease of the
energy stored in the field due to the excitation of a tachyonic mode 
$\hat c^\dagger_\Omega |0\rangle_{\rm in}$ with negative energy 
expectation value [see Eq.~(\ref{unbounded_from_below})] and corresponding 
discussion). The copious excitation of the detector realizes the fact 
that in the unstable phase the scalar field functions as an energy 
reservoir only limited by backreaction effects.

\section{ Falling asleep of the vacuum and particle creation}
\label{particle creation}

\subsection{General discussion}
\label{General_discussion}

As already mentioned, at some point the unstable phase must cease, 
leading the system back to some stationary configuration. This 
will be represented by the static region~C in Eq~(\ref{metric1}), 
which completes the scenario presented by Eq.~(\ref{metric0}):
\begin{eqnarray}
ds^2=  \left\{
 \begin{array}{ll}
                    -dt^2 + d {\bf x}^{2}   &\,\,\, {\rm (A)} \\
      N^2_{\rm (B)}(-dt^2 + h^{\rm (B)}_{i j}dx^i dx^j) &\,\,\, {\rm (B)} \\
      N^2_{\rm (C)}(-dt^2 + h^{\rm (C)}_{i j}dx^i dx^j)  &\,\,\, {\rm (C)} \\
 \end{array} 
        \right..
  \label{metric1}
 \end{eqnarray}
Here, $N_{(J)}=N_{(J)}({\bf x})>0$, $J \in \{ {\rm B, C} \}$, 
are smooth functions and $h^{(J)}_{i j}=h^{(J)}_{i j}({\bf x})$  
($i,j=1,2,3$). We note that for the sake of 
simplicity we are using the same coordinate notation $(t, {\bf x})$ for 
the three epochs. 

Because the field is assumed to be deprived of tachyonic modes in
region~C, we expand $\hat \Phi$ as 
\begin{equation}
\hat{\Phi}  =
\int d\mu (\sigma) [\hat d_{\sigma} \nu^{(+)}_{\sigma} 
+ \hat d^{\dagger}_{\sigma} \nu^{(-)}_{\sigma}], 
\label{Phi_C}
\end{equation}
where $\{ \nu_\sigma^{(+)}, \nu_\sigma^{(-)} \}$ are normal modes
which in region~C satisfy 
\begin{equation}
\nu^{\rm (+)}_{\sigma}\stackrel{{\rm (C)}}{=} 
\frac{e^{- i \varpi_{\sigma}t}}{\sqrt{2\varpi_{\sigma} \,} N_{\rm (C)}({\bf x})}
F^{\rm (C)}_{\sigma}({\bf x}), 
\;\;\; \varpi_{\sigma} > 0,
\label{outmodes}
\end{equation}
and analogously for $\nu^{\rm (-)}_{\sigma}$. All symbols in Eq.~(\ref{outmodes}) 
can be  inferred from Eq.~(\ref{intmodes}) by replacing ``B" by ``C". 
Moreover, because region~C is also static, we may wonder what will be 
the particle content of the scalar field in this region. The key point
consists in realizing that the in-vacuum fluctuations which were 
exponentially amplified during the unstable phase cannot, in general, 
be accommodated as mere fluctuations of the out-vacuum state (the one defined
by $d_\sigma |0\rangle_{\rm out} \equiv 0$ for all $\sigma$ and which represents
absence of particles according to static observers in region~C). 
In conclusion, a burst of particles is expected as the field exits the unstable 
phase~B.

Let us estimate the expectation number of created particles in 
the simplified case where the spacetime is symmetric by 
time reflection with respect to some Cauchy surface 
$\Sigma_{t_S}$ in region B. Hence, we consider a
particular case of Eq.~(\ref{metric1}), namely, 
\begin{eqnarray}
ds^2=  \left\{
 \begin{array}{ll}
                    -dt^2 + d {\bf x}^{2}   &\,\,\, {\rm (A)} \\
      N^2_{\rm (B)}(-dt^2 + h^{\rm (B)}_{i j}dx^i dx^j) &\,\,\, {\rm (B)} \\
                    -dt^2 + d {\bf x}^{2}   &\,\,\, {\rm (C)} \\
 \end{array} 
        \right.,
        \label{metric_general_discussion}
 \end{eqnarray}
although we emphasize that we have chosen regions~A and~C to be flat
only for the sake of simplicity; the same reasoning presented below 
may be straightforwardly applied to other static spacetimes.
Now, we focus on the normal modes $u^{(\pm)}_{\bf k}$ and
$\nu^{(\pm)}_{\bf k}$ with respect to which asymptotic observers 
in regions~A and~C define their no-particle states, respectively. 
In the past and future regions, they assume the following forms: 
\begin{eqnarray}\label{u_modes_star_A}
u^{(\pm)}_{\bf k}\stackrel{{\rm (A)}}{=}
(16\pi^3\omega_{\bf k})^{-1/2} \exp[\mp i(\omega_{\bf k}t - {\bf k}\cdot{\bf x})],
\\
\label{u_modes_star_C}
\nu^{(\pm)}_{\bf k}\stackrel{{\rm (C)}}{=}
(16\pi^3\omega_{\bf k})^{-1/2} \exp[\mp i(\omega_{\bf k}t - {\bf k}\cdot{\bf x})].
\end{eqnarray}
We are interested in $u_{\bf k}^{(\pm)}$ and $\nu^{(\pm)}_{\bf k}$ evolved 
forward and backward to the beginning and end of region~B, namely, 
$u_{\bf k}^{(\pm)} (t_0,{\bf x})$ and $\nu^{(\pm)}_{\bf k} (t_0+T,{\bf x})$,
respectively. Here, $t=t_0 \equiv t_S - T/2$ determines the beginning of the unstable phase~B  
and $T$ represents its coordinate-time duration. Because of our assumption 
that the spacetime is symmetric by time reflection with respect to $\Sigma_{t_S}$, 
we have (up to global phases) 
$$
u_{\bf k}^{(+)} (t_S -t,{\bf x}) 
= 
\nu^{(-)}_{- {\bf k}} (t_S + t,{\bf x}).
$$
In particular, for $t=T/2$:
$$
\nu^{(-)}_{- {\bf k}} (t_0+T,{\bf x})
\stackrel{{\rm (B)}}{=} u_{\bf k}^{(+)} (t_0,{\bf x}).
$$
Then, we use Eq.~(\ref{bog}) to decompose 
$u^{\rm (+)}_{{\bf k}} $  in terms of $v_\sigma^{(\pm)}$ 
and $w_\Omega^{(\pm)}$, obtaining
\begin{eqnarray}
\nu^{(-)}_{-{\bf k}} (t_0+T,{\bf x})
&\stackrel{{\rm (B)}}{=}  & 
\alpha^*_{\Omega {\bf k}}w^{(+)}_{\Omega} (t_0 ,{\bf x})
-  \beta_{\Omega {\bf k}}w_{\Omega}^{(-)} (t_0,{\bf x})
\nonumber \\
& + &
\int d\mu(\sigma)[\alpha^*_{\sigma {\bf k}} v_{\sigma}^{(+)} (t_0,{\bf x})
\nonumber \\
& - & \beta_{\sigma {\bf k}}v_{\sigma}^{(-)} (t_0,{\bf x})],
\label{X}
\end{eqnarray}
where  we assume again the existence of a single tachyonic mode for the sake of simplicity. 

In order to investigate particle creation in region~C, we must, e.g., project 
$u^{\rm (+)}_{{\bf k}} (t_0+T,{\bf x})$ into $\nu^{(-)}_{\bf k} (t_0+T,{\bf x})$, 
where [see Eq.~(\ref{bog})]:
\begin{eqnarray}
u^{\rm (+)}_{{\bf k}} (t_0+T,{\bf x}) 
& = &
\alpha^*_{\Omega {\bf k}}w^{(+)}_{\Omega} (t_0+T,{\bf x})
\nonumber \\
& - &\beta_{\Omega {\bf k}}w_{\Omega}^{(-)} (t_0+T,{\bf x})
\nonumber \\
& + &
\int d\mu(\sigma)[\alpha^*_{\sigma {\bf k}} v_{\sigma}^{(+)} (t_0+T,{\bf x}) 
\nonumber \\
& - & 
\beta_{\sigma {\bf k}}v_{\sigma}^{(-)} (t_0+T,{\bf x})].
\label{X2}
\end{eqnarray}
It is easy to see that  
\begin{equation}
v^{(\pm)}_{\sigma} (t_0+T,{\bf x}) = 
\exp (\mp i \varpi_\sigma T) v^{(\pm)}_{\sigma} (t_0, {\bf x}),
\label{vplusT}
\end{equation}
while we obtain from Eq.~(\ref{intmodes2}) that
\begin{eqnarray}
w^{(\pm)}_\Omega (t_0 + T,{\bf x})
&=&  \pm 2 i\sinh (\Omega T \mp i 
\pi/6)w^{(\pm)}_\Omega (t_0,{\bf x})
\nonumber \\
&&
\mp 2 i \sinh (\Omega T)
w_\Omega^{(\mp)}(t_0,{\bf x}).
\label{wplusT}
\end{eqnarray}
Then, by using Eqs.~(\ref{X}) and~(\ref{X2}), we obtain for large enough
$\Omega T$ that
$$
( \nu_{{\bf k}'}^{\rm (-)}, u^{(+)}_{{\bf k}} )_{\rm KG}
\sim e^{\Omega T} \zeta_{ {\bf k} {\bf k}'},
$$
where the Klein-Gordon inner product was realized on the $\Sigma_{t_0+T}$
Cauchy surface and
$$
\zeta_{{\bf k} {\bf k}'} = 
   i [(\alpha^*_{\Omega {\bf k}} e^{-i\pi/6} 
     - \beta_{\Omega {\bf k}}) \alpha_{\Omega -{\bf k}'}
     -(\alpha^*_{\Omega {\bf k}} 
     - \beta_{\Omega {\bf k}} e^{i\pi/6} ) \beta^*_{\Omega -{\bf k}'}].
$$
This leads to an expectation number of created particles with quantum 
numbers
${\bf k}'$ given by 
$$
\langle N_{{\bf k}'} \rangle 
\sim
e^{2 \Omega T} \int d^3 k\,  | \zeta_{{\bf k} {\bf k}'} |^2,
$$
which grows exponentially as scaled by the product $\Omega T$.
In particular, even if the transitions from regions~A to~B
and from regions~B to~C were made arbitrarily slow 
in order to minimize any particle creation due to background 
change $(\beta_{\sigma {\bf k}} \approx 0)$, 
this would not alter the fact that a large amount of particles 
would be eventually created as the vacuum falls asleep (at least 
in the present scenario; see additional comments at the end 
of Sec.~\ref{toy model}).

Next, we shall show that the burst of particles calculated above 
does not rely on phase~B being static; it will occur as long as the 
in-vacuum fluctuations get significantly amplified.

\subsection{ A toy model}
\label{toy model}

In order to illustrate our general conclusion above, let us make an explicit 
calculation assuming a concrete scenario complying with the asymptotic static 
regions~A and~C considered in Eq.~(\ref{metric_general_discussion}) but assuming 
some time evolution in the intermediate region~B. This is in agreement with the 
idealized situation where initially spread out matter collapses to form a compact object and 
eventually disperses back to infinity. Instead of calculating the particle production
over the whole space, we shall restrict attention to the interior of a small
cubical box with coordinate volume $L^3$ (oblivious to the matter forming the star), 
initially empty  (of $\Phi$ particles), placed in the very beginning at the spatial position
where the star core will form. The convenience of introducing a small box is that we 
can cover its interior with approximately Cartesian spatial coordinates 
${\bf {\tilde x}}$ (in which first-order spatial derivatives of the metric are 
negligible), writing the line element as 
\begin{equation}
 ds^2 \approx a^2 (-dt^2 + d{\bf {\tilde x}}^2).
\label{metricbox}
\end{equation}
Here, $a=a(t)> 0$ is introduced to reflect the background time evolution at the 
star's center ($a=1$ in regions~A and~C) and we have omitted the second-order 
spatial dependence of the metric (which, nevertheless, contribute 
to the scalar-curvature term). The background evolution is chosen such that 
at some point the vacuum in the box is awaken by the presence of (six-fold 
degenerate) tachyonic modes~\cite{clarification}. After the unstable phase is 
terminated, we calculate the number of massless scalar particles which
were created inside the box.

Using Eq.~(\ref{metricbox}), we write Eq.~(\ref{kg_eq}) as 
\begin{equation}
\frac{1}{a^4}\frac{\partial}{\partial t}\left(a^2 \frac{\partial \Phi}{\partial t}\right)-\frac{1}{a^2}\nabla^2\Phi+ \xi R \Phi
=0,
\label{phi_tilde}
\end{equation}
where $\nabla^2 \equiv \sum_j \partial^2/{\partial {\tilde x}^j}^2$ 
is the usual Laplace operator. Assuming, for the sake of simplicity, 
periodic boundary conditions, we look for solutions of Eq.~(\ref{phi_tilde}) 
in the form
\begin{equation}
\phi_{\bf k} (t,{\bf \tilde  x})
=
\frac{\chi_{\bf k}(t)}{a(t) \sqrt{L^3\,}} e^{ i{\bf k} \cdot {\bf \tilde  x}},
\label{phik}
\end{equation}
where  
${\bf k}\equiv 2\pi {\bf n}/L$ with
${\bf n}\in \mathbb{Z}^3$ (${\bf n}\neq {\bf 0}$). 
By using Eq.~(\ref{phik}) in Eq.~(\ref{phi_tilde}) we find 
that
\begin{equation}
\left[- \frac{d^2}{dt^2} - V_{\rm eff} (t) \right]\chi_{{\bf k}}
= {\bf k}^2 \chi_{{\bf k}},
\label{chi}
\end{equation}
where
\begin{equation}
V_{\rm eff} (t)= a^2 \xi R - a^{-1} d^2 a/dt^2.
\label{V_eff}
\end{equation}
Equation~(\ref{phik}) makes explicit another neat feature of introducing
the small box: the boundary condition which it imposes locks the
spatial dependence of the modes so that the time evolution can only
mix modes with the same ${\bf \tilde x}$ dependence. This property will
be used later to simplify the Bogoliubov-coefficient calculation.

\begin{figure}[t]
\vspace{0.3cm}
\begin{center}\includegraphics[height=0.25 \textheight]{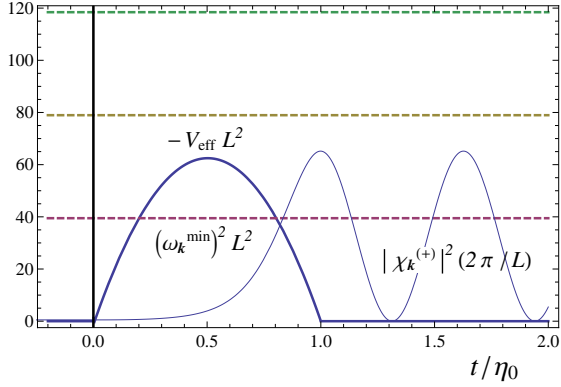}
\caption{
We plot $- V_{\rm eff}$ as a function of $t$ for $V_0  = 1.6 /(L/2 \pi)^2$ 
with the three smallest  $\omega_{\bf k } \equiv 2 \pi \| {\bf n} \|/L$ 
possible values (see horizontal dashed lines). We also plot 
$|\chi_{\bf k}^{(+)}|^2$ for $\| {\bf k} \| = \omega_{\bf k }^{\rm min}$ 
assuming $\eta_0 = 5 L/ 2\pi$. Initially $|\chi_{\bf k}^{(+)}|^2 $ 
equals $L/4\pi$, then grows exponentially in the unstable region, where 
$-V_{\rm eff} -(\omega_{\bf k }^{\rm min})^2 > 0$, and eventually oscillates 
around $(|\beta_{\bf k}|^2 + 1/2) (L/2\pi)$. The large amplitude 
which characterizes $|\chi_{\bf k}^{(+)}|^2$ at the end of the 
unstable phase reflects the fact that the in-vacuum fluctuations 
do not evolve into mere fluctuations of the out-vacuum state. } 
\label{fig1}
\end{center}
\end{figure}

Now, we assume that energy density and pressure of ordinary matter at 
the center of the star drives $R$ in Eq.~(\ref{V_eff}) to induce the 
following simple form for the effective potential:  
\begin{eqnarray}
V_{\rm eff} (t)
=  \left\{
 \begin{array}{l}
  0                                       
  \,\,\,\, {\rm for}\,\,\,\,  t \leq 0 \,\,\,\, {\rm and}\,\,\,\, t \geq \eta_0 \\
  4 V_0 (t /\eta_0) ( t/\eta_0 - 1 )
  \,\,\,\, {\rm for} \,\,\,\,  0 < t < \eta_0 
 \end{array} 
        \right.,
        \label{V_eff2}
 \end{eqnarray}
where $\eta_0, V_0 = \rm {const} > 0$.  We see that $V_{\rm eff} (t)$
has a parabolic form in the region $ 0 < t < \eta_0$
and reaches its minimum, $-V_0$, at $t = \eta_0/2$ (see Fig.~\ref{fig1}).
   
Convenient sets of in-modes $\{U^{(\pm)}_{\bf k}\}$ for $t\leq 0$ and 
out-modes $\{ V^{(\pm)}_{\bf k}\}$ for $t \geq \eta_0$ 
are exhibited below:
\begin{equation}
U^{(\pm)}_{{\bf k}} \stackrel{{t\leq 0}}{=} 
\frac{e^{\mp i(\omega_{\bf k} t- {\bf k}\cdot 
{\bf {\tilde x}} )}}{\sqrt{2 L^3 \omega_{{\bf k}}}}, 
\,\,\,\,
V^{(\pm)}_{{\bf k}} \stackrel{{t \geq \eta_0}}{=} 
\frac{e^{\mp i(\omega_{\bf k} t- {\bf k}\cdot 
{\bf {\tilde x}} )}}{\sqrt{2 L^3 \omega_{{\bf k}}}},
\label{UandW}
\end{equation}
where $\omega_{{\bf k}}\equiv \|{\bf k}\|$.
The general expression of $U^{(\pm)}_{{\bf k}}$ which complies with 
Eq.~(\ref{phik}) and fits with its form~(\ref{UandW}) in region~A  is
\begin{equation}
U^{(\pm)}_{{\bf k}}(t,{\bf {\tilde x}})=
\frac{\chi^{(\pm)}_{{\bf k}}(t)}{a(t)\sqrt{L^3}}  
e^{\pm i{\bf k}\cdot {\bf {\tilde x}}} 
\label{U_general}
\end{equation}
with
$
\chi^{(\pm)}_{{\bf k}}(t\leq0) =
{e^{\mp i\omega_{{\bf k}}t}}/{\sqrt{2\omega_{\bf k}}}.
$ 
From the spatial dependence of the modes, we readily see that 
[recall the discussion below Eq.~(\ref{V_eff})]
\begin{equation}
U^{(+)}_{\bf k}(t, {\bf \tilde x}) = 
\alpha_{\bf k} V^{(+)}_{\bf k}(t, {\bf \tilde x})
+
\beta_{- {\bf k}} V^{(-)}_{- {\bf k}} (t, {\bf \tilde x}),
\label{jogo da velha}
\end{equation}
where the Bogoliubov coefficients between the bases $\{U^{(\pm)}_{\bf k}\}$
and $\{ V^{(\pm)}_{\bf k'}\}$ are 
$$
\alpha_{\bf k k'} = \alpha_{\bf k'} \delta_{\bf k k'},\,\,\,
\beta_{\bf k k'} = \beta_{\bf k'} \delta_{{\bf k} \,-{\bf k'}}.
$$
For $\chi^{(+)}_{\bf k}$ in region~C, Eqs.~(\ref{UandW}), (\ref{U_general}), 
and~(\ref{jogo da velha}) imply
$$
\chi^{(+)}_{\bf k} (t) \stackrel{{t\geq \eta_0}}{=}
\alpha_{\bf k} \frac{e^{-i \omega_{\bf k} t}}{\sqrt{2\omega_{\bf k}}}
+
\beta_{-{\bf k}} \frac{e^{ i \omega_{\bf k} t}}{\sqrt{2\omega_{\bf k}}},
$$
from where $\alpha_{\bf k} $ and $\beta_{\bf k} = \beta_{-{\bf k}}$ 
can be easily obtained in terms of  $\chi^{(+)}_{\bf k}$ and 
${\dot \chi}^{(+)}_{\bf k} \equiv d\chi^{(+)}_{\bf k}/dt$
evolved into region~C:
\begin{eqnarray}
\alpha_{\bf k} 
& = &
\left[ \frac{e^{i \omega_{\bf k} t}}{\sqrt{2 \omega_{\bf k}}} 
\left(\omega_{\bf k} \chi^{(+)}_{\bf k} + i {\dot \chi}^{(+)}_{\bf k}\right)
\right]_{t\geq \eta_0},
\nonumber \\
\beta_{\bf k} 
& = &
\left[ \frac{e^{-i \omega_{\bf k} t}}{\sqrt{2 \omega_{\bf k}}} 
\left(\omega_{\bf k} \chi^{(+)}_{\bf k} - i {\dot \chi}^{(+)}_{\bf k}\right)
\right]_{t\geq \eta_0}.
\nonumber
\end{eqnarray}
(It can be easily verified using Eqs.~(\ref{chi}) and~(\ref{V_eff2}) that the
expressions for $\alpha_{\bf k}$ and~$\beta_{\bf k}$ above do not depend on
the value of $t\geq \eta_0$.)  
Therefore, assuming that the field is initially in the no-particle state 
$|0\rangle_{\rm in}$ with respect to asymptotic past observers 
as defined by the in-modes $U^{(\pm)}_{{\bf k}}$, the expectation value of created 
particles  in region~C~\cite{BD82, Fulling89}
$$
_{\rm in}\langle 0|\hat{N}_{\rm out}|0 \rangle_{\rm in}
= 
\sum_{{\bf k},{\bf k}'} |\beta_{{\bf k} {\bf k}'}|^2
$$
is given by
\begin{eqnarray}
_{\rm in}\langle 0|\hat{N}_{\rm out}|0 \rangle_{\rm in}
& = &
\sum_{{\bf k}} |\beta_{{\bf k}}|^2
\nonumber \\
& = &
\sum_{\bf k} 
\left[ 
\frac{|{\dot \chi}^{(+)}_{\bf k}|^2}{2 \omega_{\bf k}}
+
\frac{\omega_{\bf k} |\chi^{(+)}_{\bf k}|^2}{2 }
-\frac{1}{2}
\right]_{t\geq \eta_0}
\label{N}
\end{eqnarray}
where 
$
\hat{N}_{\rm out}\equiv 
\sum_{\bf k}
\hat{d}^{{\rm out} \dagger}_{{\bf k}} 
\hat{d}^{\rm out}_{{\bf k}}
$
with 
$\hat{d}^{\rm out}_{{\bf k}}$ 
and 
$\hat{d}^{{\rm out} \dagger}_{{\bf k}}$ 
being the annihilation and creation operators defined with respect 
to the out-modes $V^{(\pm)}_{{\bf k}}$, respectively.
This provides an expression for obtaining the expectation
number $|\beta_{\bf k}|^2$ of created particles with quantum
numbers ${\bf k}$ once the oscillatory in-mode $\chi^{(+)}_{\bf k}$
is (numerically) evolved until region~C.

\begin{figure}[t]
\vspace{-0.4cm}
\begin{center}\includegraphics[height=0.28\textheight]{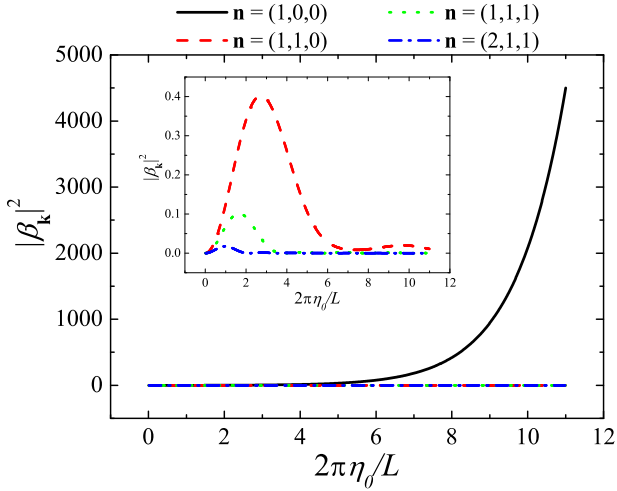}
\caption{The  expectation value of created particles $|\beta_{\bf k}|^2$ 
with quantum numbers ${\bf k} = 2\pi {\bf n}/L$ is exhibited as a 
function of $\eta_0$, where we have assumed that for some 
time interval the star becomes  dense enough 
such that $- V_{\rm eff} - ({\omega_{\bf k}^{\rm min}})^2 >0$,
while for $\omega_{\bf k} > \omega^{\rm min}_{\bf k}$ we
always have $- V_{\rm eff} - \omega_{\bf k}^2 <0$  (see 
Fig.~\ref{fig1}). Here, we have chosen $V_0 (L/2 \pi)^2 = 1.6 $.} 
\label{fig2}
\end{center}
\end{figure}

The assumption to illustrate the appearance of tachyonic modes consists 
in choosing a star which becomes dense enough and a coupling $\xi$ such 
that $- V_{\rm eff} - (\omega^{\rm min}_{\bf k})^2>0$ 
for the least energetic (six-fold degenerate) modes allowed in the 
box, $\omega_{\bf k}^{\rm min} = 2\pi/L$, for some time 
interval. As a result, the corresponding  $\chi_{\bf k}^{(+)}$ 
solutions  satisfying Eq.~(\ref{chi}) are verified to exponentially
grow for some time rather than to oscillate, triggering the 
vacuum awakening effect (see Fig.~\ref{fig1}). 
In Fig.~\ref{fig2}, we plot $|\beta_{{\bf k}}|^2$ as 
a function of $\eta_0$, which scales with the time interval during 
which the vacuum stays awakened. 

Clearly, the final state is dominated by modes with 
$\omega^{\rm min}_{\bf k} = 2\pi/L$, which have experienced a 
phase of exponential growth. The intensity of the particle burst 
is strongly influenced by how long the vacuum remains awake. The inset 
of Fig.~\ref{fig2} focuses on modes with $\omega_{\bf k} > 2\pi/L$, which are not 
exponentially enhanced, and stresses the usually modest
particle creation observed in time-varying spacetimes with 
asymptotic flat regions~\cite{BD82,Fulling89}. 
We note that in the adiabatic limit, where the background 
geometry changes arbitrarily slowly ($\eta_0 \to \infty$),
particle creation for  modes with $\omega_{\bf k} > 2\pi/L$
goes to zero as expected, in contrast to the ones for 
$\omega^{\rm min}_{\bf k} = 2\pi/L$,
which diverges. For a $10~{\rm m}$ 
side box, an awakening time interval corresponding to
$\eta_0 \sim 10^{-6}~{\rm s}$ would eventually lead to a massive 
creation of particles, with energy $2 \pi/L$, engendering densities 
of $10^{14}~{\rm g/cm}^3$, which is the typical density for some compact 
stars. If we relax our small box assumption and take 
$ L \sim 10~{\rm km}$, the same density 
would be reached for $\eta_0 \sim 10^{-3}~{\rm s}$. Interestingly 
enough, this corresponds to the time interval for the vacuum energy 
density to take control over the evolution of the compact star 
once the vacuum awakening effect is triggered [see discussion below 
Eq.~(\ref{estimativa})]. 

We stress that the conclusions above are derived by assuming  
the effective potential~(\ref{V_eff2}), which is symmetric by 
time reflection, and can change depending on the  
star evolution. In order to show this, let us discuss the
energetics of particle creation in the context of our toy model.
For this purpose, it is useful to make the transformation 
$\Phi \to {\tilde \Phi} =a \Phi$
to write Eq.~(\ref{phi_tilde}) for ${\tilde \Phi}$ as 
\begin{equation}
\left[
{\partial^2}/{\partial t^2}-\nabla^2+ V_{\rm eff}(t)
\right] 
\tilde{\Phi}=0.
\label{phi_tilde3}
\end{equation}
Thus, we have translated the problem into the simpler one of a 
scalar field  ${\tilde \Phi}$ in a flat spacetime 
$(\mathbb{R}^4,\eta_{ab})$ subject to an external 
time-dependent potential $V_{\rm eff}(t)$. 

The action which gives rise to Eq.~(\ref{phi_tilde3}) and its corresponding 
stress-energy tensor are 
\begin{equation}
S \equiv 
-\frac{1}{2}
\int_{\mathbb{R}^4} d^4x \sqrt{-\eta} \,
( \partial_a\tilde{\Phi} \partial^a\tilde{\Phi}^* + V_{\rm eff}\tilde{\Phi}
\tilde{\Phi}^* )
\label{action_toy}
\end{equation} 
and
\begin{equation}
T_{ab} 
= 
\partial_{(a}\tilde{\Phi}\partial_{b)}\tilde{\Phi}^* 
- \frac{1}{2}\eta_{ab}
\left[
\partial_c\tilde{\Phi}\partial^c\tilde{\Phi}^*
+ V_{\rm eff}\tilde{\Phi}\tilde{\Phi}^*
\right],
\label{stressenergy_toy}
\end{equation}
respectively.
Next, it can be shown from Eqs.~(\ref{phi_tilde3}) and~(\ref{stressenergy_toy}) 
that
\begin{equation}
\partial_a\left(T^a_{\; b}(\partial_t)^b\right)=-\frac{1}{2}
\frac{d V_{\rm eff}}{dt} |\tilde{\Phi}|^2,
\label{divT}
\end{equation}
which can be rewritten as 
\begin{equation}
\frac{\partial \rho}{\partial t}+\nabla \cdot {\bf j}
=
\frac{1}{2}\frac{d V_{\rm eff}}{d t}|\tilde{\Phi}|^2,
\label{energy_cons}
\end{equation}
where 
\begin{equation}
\rho \equiv 
\frac{1}{2}\left(\frac{\partial \tilde{\Phi}}{\partial t}\frac{\partial \tilde{\Phi}^*}{\partial t}
+ 
\nabla \tilde{\Phi}\cdot \nabla \tilde{\Phi}^*
+ V_{\rm eff}|\tilde{\Phi}|^2 \right)
\nonumber
\end{equation} 
and 
\begin{eqnarray*}
{\bf j} 
\equiv   
-\frac{1}{2}\left(\frac{\partial\tilde{\Phi}^*}{\partial t}
\nabla \tilde{\Phi}
+  
\frac{\partial \tilde{\Phi}}{\partial t} \nabla \tilde{\Phi}^*\right).
\end{eqnarray*} 

We see from Eq.~(\ref{energy_cons}) that the energy stored in the scalar 
field is not locally conserved whenever $dV_{\rm eff}/dt \neq 0$. The 
extra energy pumped into or out of 
the field is accounted by the ``external agent" responsible to change 
$V_{\rm eff}$. Moreover, notice from Eq.~(\ref{energy_cons})
that even if $V_{\rm eff} (t)$ is symmetric under time reflection,
the decrease in the field energy when it enters the unstable 
phase ($d V_{\rm eff}/dt<0$ and small vacuum fluctuations 
$\langle {\hat \Phi}^2 \rangle$) is more than compensated by the increase
in the field energy when it exits the unstable phase ($d V_{\rm eff}/dt>0$ and
large vacuum fluctuations $\langle {\hat \Phi}^2 \rangle$). 
In fact, the latter can be overwhelmingly larger than
the former, with the net extra energy being responsible for the particle burst. 
This analysis implies that in a physical situation, the 
final verdict concerning the amount of created particles will 
depend on a more detailed understanding on the spacetime evolution in 
the unstable phase, which would inform us about how long the vacuum 
would stay awake, and on the final classical configuration 
reached by the gravitational and scalar fields, which would tell us how much 
energy would turn out available for particle creation.  

A closing remark for this section is in order. 
In our calculations the expectation value 
of the field remains zero
throughout the background evolution. However, this field configuration
$\langle \hat\Phi \rangle=0$ is
obviously unstable during the intermediate phase when the vacuum is awake.
Thus, one may speculate whether during the transition to the intermediate 
phase the classical field profile $\langle \hat\Phi \rangle$ could continuously 
adjust itself to nonzero values (``continuous spontaneous scalarization'') in order to 
stabilize the system, in which case tachyonic-like modes would never really
be
present. Unfortunately, a definite verdict to this question is beyond the scope of
semiclassical gravity since it involves the subtleties of decoherence of a {\it free}
field, initially in a state which is symmetric by the exchange $\Phi \leftrightarrow -\Phi$, to
a symmetry-broken phase in which $\langle \hat\Phi \rangle\neq 0$. Notwithstanding, 
a reasonable conjecture seems to be that coherence 
can be sustained for as long as the background geometry
(which can be regarded as the {\it sole} classical ``apparatus'' with which
the field interacts) is oblivious to $\Phi$. In other words, it seems quite possible
that 
$\langle \hat\Phi \rangle=0$ until backreaction
becomes important. But when that happens, the fluctuations
$\langle \hat\Phi^2 \rangle$ will already be amplified to the point where
they cannot be accommodated as mere vacuum fluctuations. A burst of particles should follow,
regardless whether $\langle \hat\Phi \rangle$ remains null or spontaneous scalarization
takes place. Another important point is that spontaneous scalarization takes place 
only for negative values of $\xi$~\cite{PCBRS11}. Therefore, for $\xi>0$ the whole scenario
of a gradually changing $\langle \hat\Phi \rangle$ ensuring the stability of the system 
seems even more
unlikely.



\section{Final remarks} 
\label{final remarks}

After a review of the vacuum awakening effect in relativistic stars,
we have probed the exponential increase of the quantum field 
fluctuations using Unruh-DeWitt detectors. The fast increase of these 
fluctuations may lead 
eventually  to an important burst of free-field particles after the 
vacuum is forced to fall asleep again. This burst of particles would 
draw a significant amount of energy from the initial system. The amount of 
created particles will depend on the duration of the unstable epoch 
and on the final gravitational and scalar field configuration,   
which are open issues at this point. A possible signal favoring the 
vacuum awakening effect for a free field would be the unveiling of 
astrophysical events outpouring less amounts of {\it visible} energy 
than would be expected. This may also provide an efficient way of 
converting  energy initially stored in the form of ordinary matter 
(forming the star) into a ``dark'' component which couples only to 
gravity. 

Simultaneous to the completion of this article, it was posted 
a classical analysis showing that during the 
scalarization process a strong emission of scalar radiation should
occur~\cite{RDANS12}, which is in line with the conclusions 
presented here, especially with the discussion at the end of the previous 
section.

\acknowledgments
The authors are grateful to J. Montero and V. Pleitez for discussions
concerning possible Planck-scale contributions to the scalar mass.
A.L., W.L. and G.M., D.V. acknowledge full and partial support from 
Fun\-da\-\c{c}\~ao de Am\-pa\-ro \`a Pes\-qui\-sa do Es\-ta\-do de  
S\~ao Paulo, respectively. G.M. also acknowledges partial support 
from Conselho Nacional de Desenvolvimento Cient\'\i fico e 
Tecnol\'ogico.


\begin{thebibliography}{}
\bibitem{errata}
We note that Eq.~(3) in Ref.~\cite{LMV10} contains an incorrect
$f(\chi)$ factor and should read as shown in Eq.~(\ref{intmodes_estrela}) 
in this article. No results in Ref.~\cite{LMV10} are affected  by
this misprint.
\bibitem{clarification}
The vacuum inside the small box is harder to awaken because its 
size introduces an ``effective mass" of order $1/(aL)$, which would spoil
the effect for realistic stars and $\xi \sim 1$. However, since this is 
a toy model, we can consider $|\xi | \gg 1$ for the sake of our illustration.
\end{thebibliography}
\end{document}